\documentclass[superscriptaddress,reprint,amsmath,twocolumn,amssymb,prx, nodate, longbibliography]{revtex4-2}

\usepackage{graphicx} 
\usepackage{amsmath} 
\usepackage{xifthen}
\usepackage{physics}
\usepackage[caption=false]{subfig}
\usepackage{subfiles}

\usepackage{floatrow}
\newfloatcommand{capbtabbox}{table}[][\Xhsize]

\usepackage{hyperref} 
\usepackage[noabbrev]{cleveref} 

\crefname{appendix}{appendix}{appendices}
\Crefname{appendix}{Appendix}{Appendices}
\usepackage{etoolbox} 
\makeatletter
\appto{\appendix}{%
  \@ifstar{\def\theequation@prefix{A.}}%
          {}%
}
\makeatother

\usepackage[load=prefixed,load=abbr,separate-uncertainty=true]{siunitx} 

\newcommand{\sub}[1]{_{\mathrm{#1}}}

\newcommand{\ie}{i.e.\@}
\newcommand{\eg}{e.g.\@}

\newcommand{\Loss}{\mathcal{L}}
\newcommand{\OD}{\alpha_0}

\newcommand{\etal}{\emph{et~al.}}

\newcommand{\nn}[2]{^{(#1)}_{#2}}

\usepackage{mathtools,eqparbox}

\crefname{figure}{Fig.}{Fig.}
\Crefname{figure}{Fig.}{Fig.}
\crefname{equation}{Eq.}{Eq.}
\Crefname{equation}{Equation}{Equation}

\usepackage{todonotes}
\usepackage[normalem]{ulem}

\begin{document}

\title{Backpropagation through nonlinear units for all-optical training of neural networks}

\author{Xianxin Guo}
\thanks{These authors contributed equally to this work.}
\email{xianxin.guo@physics.ox.ac.uk}
\affiliation{Institute of Fundamental and Frontier Sciences, University of Electronic Science and Technology of China, Chengdu, Sichuan 610054, China}
\affiliation{University of Oxford, Clarendon Laboratory, Parks Road, Oxford  OX1 3PU, UK}
\affiliation{Institute for Quantum Science and Technology, University of Calgary, Calgary, Canada, T2N 1N4}

\author{Thomas D.\ Barrett}
\thanks{These authors contributed equally to this work.}
\email{thomas.barrett@physics.ox.ac.uk}
\affiliation{University of Oxford, Clarendon Laboratory, Parks Road, Oxford  OX1 3PU, UK}

\author{Zhiming M.\ Wang}
\email{zhmwang@uestc.edu.cn}
\affiliation{Institute of Fundamental and Frontier Sciences, University of Electronic Science and Technology of China, Chengdu, Sichuan 610054, China}

\author{A.\ I.\ Lvovsky}
\email{alex.lvovsky@physics.ox.ac.uk}
\affiliation{University of Oxford, Clarendon Laboratory, Parks Road, Oxford  OX1 3PU, UK}
\affiliation{Russian Quantum Center, Skolkovo, 143025, Moscow, Russia}

\begin{abstract}

Backpropagation through nonlinear neurons is an outstanding challenge to the field of optical neural networks and the major conceptual barrier to all-optical training schemes.  Each neuron is required to exhibit a directionally dependent response to propagating optical signals, with the backwards response conditioned on the forward signal, which is highly non-trivial to implement optically.  We propose a practical and surprisingly simple solution that uses saturable absorption to provide the network nonlinearity.  We find that the backward propagating gradients required to train the network can be approximated in a pump-probe scheme that requires only passive optical elements.  Simulations show that, with readily obtainable optical depths, our approach can achieve equivalent performance to state-of-the-art computational networks on image classification benchmarks, even in deep networks with multiple sequential gradient approximations.  This scheme is compatible with leading optical neural network proposals and therefore provides a feasible path towards end-to-end optical training.

\end{abstract}

\maketitle

\section{Introduction}

Machine learning (ML) is changing the way in which we approach complex tasks, with applications ranging from natural language processing~\cite{cambria14} and image recognition~\cite{rawat17} to artificial intelligence~\cite{silver16} and fundamental science~\cite{gilmer17,torlai18}. At the heart (or `brain') of this revolution are artificial neural networks (ANNs), which are universal function approximators~\cite{hornik89, cybenko89} capable, in principle, of representing an arbitrary mapping of inputs to outputs. Remarkably, their function only requires two basic operations: matrix multiplication to communicate information between layers, and some non-linear transformation of individual neuron states (activation function). The former accounts for most of the computational cost associated with ML.
This operation can, however, be readily implemented by leveraging the coherence and superposition properties of linear optics~\cite{tamura79}. Optics is therefore an attractive platform for realising the next generation of neural networks, promising faster computation with low power consumption~\cite{shen17, deMarinis19}.

Proposals for optical neural networks (ONNs) have been around for over thirty years~\cite{abu87, jutamulia96}, and have been realised in both free-space~\cite{bueno18, lin18, zuo19} and integrated~\cite{shen17} settings. However, the true power of neural networks is not only that they can approximate arbitrary functions, but also  that they can ``learn'' that approximation.  The training of neural networks is, almost universally, achieved by the backpropagation algorithm \cite{lecun15}. Implementing this algorithm  optically is challenging because it requires the the response of the network's nonlinear elements to be different for light propagating forwards or backwards. Confronted with these challenges, existing ONNs are actually trained with, or heavily aided by, digital computers~\cite{shen17,bueno18,zuo19,hughes18}. As a result, the great advantages offered by optics remain largely unexploited. Developing an all-optically trained ONN to leverage these advantages remains an unsolved problem.  Here, we address this challenge and present a practical training method capable of backpropagating the error signal through nonlinear neurons in a single optical pass.

The backpropagation algorithm aims to minimise a loss function that quantifies the divergence of the network's current performance from the ideal, via gradient descent~\cite{lecun15}.  To do so, the following steps are repeated until convergence: (1) forward propagation of information through the network; (2) evaluation of the loss function gradients with respect to the network parameters at the output layer; (3) backpropagation of these gradients to all previous layers; (4) parameter updates in the direction that maximally reduces the loss function.  Forward propagation (step (1)) requires both the aforementioned matrix multiplication, which maps information between layers,  and a suitable nonlinear activation function, which is applied individually to each neuron.  Whilst this nonlinearity has so far been mostly applied digitally in hybrid optical-electronic systems~\cite{shen17, hughes18, williamson19} -- at the cost of repeatedly measuring and generating the optical state -- recent work has also realised optical nonlinearites~\cite{cheng13, zuo19}.

However, obtaining and backpropagating the loss-function gradients (steps (2-3)) remains an outstanding problem in an optical setting.  Whilst backpropagating through the linear interconnection between layers is rather straightforward, as linear optical operations are naturally bidirectional, the nonlinearity of neurons is a challenge.  This is because the backwards-propagating signal must be modulated by the derivatives of the activation function of each neuron at its current input value, and these derivatives are not readily available in an ONN.

In 1987, Wagner~\etal{} suggested that a feedforward ONN could be implemented and trained by using Fabry-Perot etalons to approximate the required forwards and backwards response of a sigmoid nonlinearity~\cite{wagner87}.  However, this backpropagation approach was never realised, or even analyzed in detail, largely due to its inherent experimental complexity, with a subsequent ONN demonstration instead using digitally calculated errors~\cite{psaltis88}.
A further approach to an optically-trained feedforward network was proposed by Cruz-Cabrera~\etal{}~\cite{cruz00}. They used  a highly non-standard network architecture that transforms a ``continuum of neurons'' (a wavefront) as it passes through a nonlinear crystal using cross-phase modulation with a secondary ``weight'' beam. In a proof-of-concept experiment, the learning of two-bit logic was demonstrated.

An additional challenge is to map from the gradients with respect to the (platform-agnostic) weight matrices to the physical parameters that control these matrices in a specific ONN platform.  In 2018, Hughes~\etal{}~\cite{hughes18} proposed an elegant method to directly obtain the gradients of these control parameters by an additional forward-propagating step.  However, this scheme assumes computing the derivatives of the activation functions digitally and applying them to the backpropagating signal electro-optically.  An extensive review of these and other related works can be found in the Supplemental Material.

This work directly addresses the issue of optical backpropagation through nonlinear units in a manner that is both consistent with modern neural network architectures and compatible with leading ONN proposals.
We consider an optical nonlinearity based on saturable absorption (SA) and show that, with the forward-propagating features and the backward-propagating errors taking the roles of pump and probe respectively, backpropagation can be realised using only passive optical elements.  Our method is effective and surprisingly simple -- with the required optical operations for both forwards and backwards propagation realised using the same physical elements.  Simulations with physically realistic parameters show that the proposed scheme can train networks to performance levels equivalent to state-of-the-art ANNs. When combined with optical calculation of the error term at the output layer via interference, this presents a path to the all-optical training of ONNs.

\section{Implementing optical backpropagation}

We begin by recapping the operation of a neural network before discussing optical implementations.  Seeded with data at the input layer ($a\nn{0}{}$), forward-propagation maps the neuron activations from layer $l{-}1$ to the neuron inputs at layer $l$ as
\begin{equation}\label{eq:forward}
z\nn{l}{j} = \sum_{i} w\nn{l}{ji} a\nn{l-1}{i}
\end{equation}
via a weight matrix $w\nn{l}{}$,  before applying a nonlinear activation function individually to each neuron, $a\nn{l}{j} = g(z\nn{l}{j})$ (with subscripts labelling individual neurons).

\begin{figure}[t]
\includegraphics[width=\textwidth]{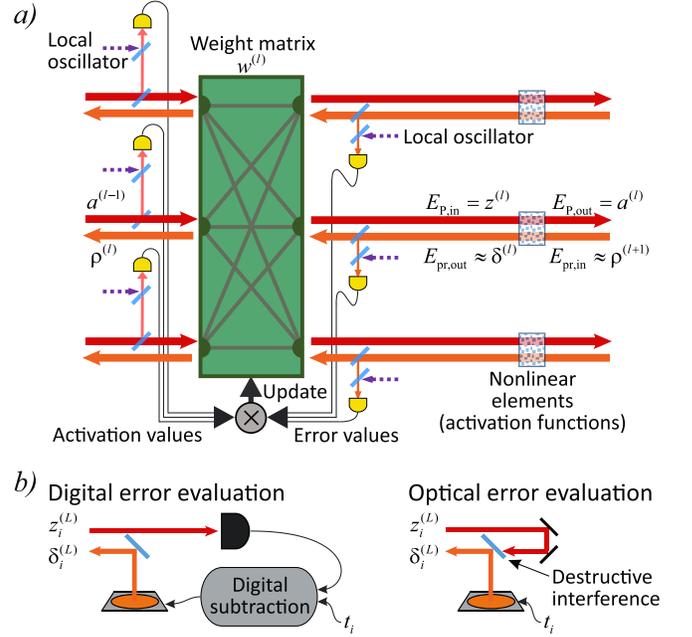}
\caption{\textbf{ONN with all-optical forward- and backward-propagation.}
\textbf{a,} A single ONN layer which consists of weighted interconnections and a SA nonlinear activation function.  The forward- (red) and backward-propagating (orange dashed) optical signals are tapped off by beam splitters and measured, using reference beams, $E\sub{ref}$, to find the neuron activations, $a\nn{l}{}$, and errors, $\delta\nn{l}{}$.  \textbf{b,} Error calculation at the output layer performed optically or digitally as described in the main text.
}
\label{fig:1}
\end{figure}

At the output layer we evaluate the loss function, $\Loss$, and calculate its gradient with respect to the weights,
\begin{equation}
\frac{\partial \Loss}{\partial w\nn{l}{ji}}
= \frac{\partial \Loss}{\partial z\nn{l}{j}}\frac{\partial z\nn{l}{j}}{\partial w\nn{l}{ji}}
= \delta\nn{l}{j}a\nn{l-1}{i},
\label{eq:weight_grad}
\end{equation}
where $\delta\nn{l}{j} \equiv \partial \Loss/\partial z\nn{l}{j}$ is commonly referred to as the `error' at the $j$-th neuron in the $l$-th layer.  From the chain rule we have
\begin{equation}
\delta\nn{l}{j}
= \sum_{k}\frac{\partial \Loss}{\partial z\nn{l+1}{k}}\frac{\partial z\nn{l+1}{k}}{\partial z\nn{l}{j}}
= g'(z\nn{l}{j})\rho_j^{(l+1)},
\label{eq:neuron_grad}
\end{equation}
where $\rho_j^{(l+1)}{=}\sum_{k}\delta\nn{l+1}{k}w\nn{l+1}{kj}$. Given the error at the output layer, \ie{} $\delta\nn{L}{}$, which is calculated directly from the loss function,  the errors $\delta\nn{L-1}{},\ldots,\delta\nn{1}{}$ for all preceding layers are sequentially found using Eq.~\eqref{eq:neuron_grad}. These errors, as well as the activations $a\nn{l-1}{}$ of all neurons, allow one to find the gradients \eqref{eq:weight_grad} of the error function with respect to all the weights, and hence apply gradient descent.

The transformation \eqref{eq:forward} is readily implemented as a linear optical (interferomentric) operation, with the neurons represented by real-valued field amplitudes in different spatial modes~\cite{tamura79}. Remarkably, calculating $\rho\nn{l+1}{}$ in the right-hand side of the backpropagation equation \eqref{eq:neuron_grad} involves the same weight matrix, meaning that it can be implemented by physical backward propagation of an optical signal through the same linear optical arrangement \cite{psaltis88}, as shown in \cref{fig:1}(a). However, multiplying this signal by the derivative of the activation function, $g'(z\nn{l}{})$, is a challenge without invoking digital electronics.

To address this challenge, we require an optical implementation of the activation function with the following features: (\romannumeral 1)~nonlinear response for the forward input; (\romannumeral 2)~linear response for the backward input; (\romannumeral 3)~modulation of backward input with the derivative of the nonlinear function. While it is natural to use nonlinear optics for this purpose, it is difficult to satisfy the requirement that the unit must respond differently to forward- and backward-propagating light.  Here, we show that this problem can be addressed using saturable absorption in the well-known pump-probe configuration.

\begin{figure}[t]
\includegraphics[width=\textwidth]{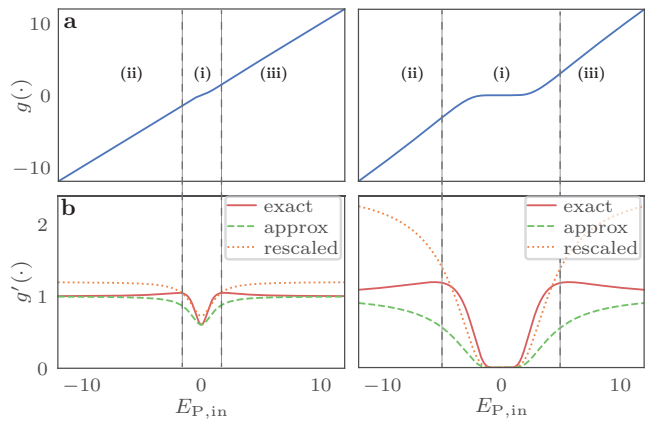}
\caption{\textbf{Saturable absorber response.}
The transmission \textbf{(a)} and the transmission derivative \textbf{(b)} of a SA unit with optical depths of 1 (left) and 30 (right), as defined by Eqs.~\eqref{eq:sat_abs_forward0} and \eqref{eq:SA_grad_exact}, respectively. Also shown in \textbf{(b)} are the actual probe transmissions given by Eq.~\eqref{eq:sat_abs_forward} which approximate the derivatives, with and without the rescaling. The scaling factors  are 1.2 (left) and 2.5 (right). Region (i) is the unsaturated (nonlinear) region exhibiting strong nonlinearity, and region (ii) is the saturated (linear) region.
}
\label{fig:SA}
\end{figure}

Consider passing a strong pump, $E\sub{P}$, and a weak probe, $E\sub{pr}$, through a two-level medium (\eg{} atomic vapour).  The pump transmission is then a nonlinear function of the input,
\begin{equation}
E\sub{P,out}=g(E\sub{P,in})  = \exp(-\frac{\OD/2}{1+E\sub{P,in}^2})E\sub{P,in},
\label{eq:sat_abs_forward0}
\end{equation}
where $\OD$ is the resonant optical depth and all fields are assumed to be normalised by the saturation threshold. \cref{fig:SA}(a) plots the pump transmission $g(\cdot)$ at $\OD$ of 1 and 30. High optical depth induces strong nonlinearity in the unsaturated region, and a sufficiently strong pump renders the medium nearly transparent in the saturated region.
A suitably weak probe, on the other hand, does not modify the transmissivity of the atomic media, and hence experiences linear absorption with the absorption coefficient determined by the pump,
\begin{equation}
\frac{E\sub{pr,out}}{E\sub{pr,in}}=\exp(-\frac{\OD/2}{1+E\sub{P,in}^2}).
\label{eq:sat_abs_forward}
\end{equation}
Note that both beams are assumed to be resonant with the atomic transition and so, as the phase of the electric field is unchanged, we treat these as real-valued without a loss of generality.  Therefore, with the pump and probe taking the roles of forward-propagating signal and backward-propagating error in an ONN, required features (\romannumeral 1) and (\romannumeral 2) of our optical nonlinear unit are met.

Condition (\romannumeral 3), however, remains to be satisfied.  The derivative of the pump transmission is
\begin{equation}
g'(E\sub{P,in})
=\left[
1 {+} \frac{\OD{E}\sub{P,in}^2}{\left(1+E\sub{P,in}^2\right)^2}
\right]
\exp(
{-}\frac{\OD/2}{1+E\sub{P,in}^2}
).
\label{eq:SA_grad_exact}
\end{equation}
The derivatives at $\OD$ of 1 and 30 are plotted in \cref{fig:SA}(b). Our key insight is that in many instances the square-bracketed factor in  Eq.~\eqref{eq:SA_grad_exact} can be considered constant, in which case, the backpropagation transmission of \eqref{eq:sat_abs_forward} is a good approximation of the desired response \eqref{eq:SA_grad_exact} up to a constant factor. Feature (\romannumeral 3) is then satisfied because a constant scaling of the network gradients can be absorbed into the learning rate.  This may appear a coarse approximation, however, as we will see in the next section, it is only required to hold within the nonlinear region of the SA response, which is the case for our system [\cref{fig:SA}(b)].

The proposed scheme can be implemented on either integrated or free-space platforms.  In the integrated setting, optical interference units that combine integrated phase-shifters and attenuators to realise intra-layer weights have been demonstrated~\cite{shen17} as has, separately, on-chip SA through atomic vapour~\cite{yang07, ritter15} and other nonlinear media~\cite{bao09, cheng13}.  A free-space implementation of the required matrix multiplication can be achieved using a spatial light modulator (SLM)~\cite{tamura79} with the nonlinear unit provided by a standard atomic vapour cell.  In the integrated case, an additional nontrivial step to map the weight gradients \eqref{eq:weight_grad} to suitable updates of the control parameters (\ie{} phase-shifters and attenuator) is required, however this challenge was recently addressed by Hughes~\etal{}~\cite{hughes18}.  A free-space implementation, by contrast, has discrete blocks of SLM pixels directly control individual weights, so the update calculation is more straightforward (see Supplemental Material for details).

Regardless of the chosen platform, passive optical elements can only implement weighted connections that satisfy conservation of energy. For networks with a single layer of nonlinear activations, this is not a practical limitation as the weight matrices can be effectively realised with normalised weights by correspondingly rescaling the neuron activations in the input layer.
For deep networks with multiple layers, absorption through the vapour cell will reduce the field amplitude available to subsequent layers.  This can be counteracted by inter-layer amplification using, for example, semiconductor optical amplifiers~\cite{connelly07}.

In our proposed ONN, the only parts that require electronics are (a)~real-valued homo- or heterodyne measurements of the tapped-off neuron activations ($a\nn{l}{}$) and error terms ($\delta\nn{l}{}$) at each layer, (b)~generating the network input and reference beams and (c)~updating the weights.  In practice, the update (c) is calculated not for each individual training set element, but as average for multiple elements (a ``mini-batch"), hence the speed of this operation is not critical for the ONN performance. Generating the inputs and targets is decoupled from the calculation performed by the ONN and requires fast optical modulators, which are abundant on the market.

Finally, the measurements (a) must be followed by calculating the product $\delta\nn{l}{j}a\nn{l-1}{i}$ and averaging over the minibatch. This operation can be implemented using electronic gate arrays. For a network with $L$ layers of $N$ neurons, this requires $2LN$ measurements and $LN^{2}$ offline multiplications. Alternatively, the multiplication can be realised by direct optical interference of the two signals with each other, followed by intensity measurement. The optical multiplication would require phase stability between the forward- and backward-propagating beams and  the additional overhead of $2LN^{2}$ photodetectors, but eliminate the need for reference beams and offline multiplications.

The primary latencies associated with the optical propagation of the signal in the ONN are due to the bandwidths of the SAs and intra-layer amplifiers. Further processing speed limitations are present in the photodetection and multiplication of $\delta\nn{l}{j}a\nn{l-1}{i}$ as well as conversion of the computed weight matrix gradients to their actuators within the ONN~\cite{hughes18}. This latter conversion however occurs once per training batch, so this limitation can be amortised by using large batches.

The remaining, not yet discussed, element of the ONN training is the calculation and re-injection of the error $\delta\nn{L}{}$ at the output layer, to initiate backpropagation. To implement this optically, we train the ONN with the mean-squared-error loss function,
\begin{equation}
	\Loss = \sum_i \tfrac{1}{2} \left( z^{(L)}_i - t_i \right)^2,
\label{eq:optical_loss}
\end{equation}
where $t_i$ is the target value for the $i$-th output neuron. This loss function implies $\delta\nn{L}{i} =  \partial \Loss/\partial z\nn{L}{i}= z\nn{L}{i} - t_{i}$, which is calculable by interference of the network outputs with the target outputs on a balanced beam-splitter.  This approach to error calculation is illustrated in the right panel of \cref{fig:1}(b), whereas the left panel shows the standard approach in which the errors are calculated offline (electronically).

\section{Examining approximation errors}

\begin{figure*}[t]
\includegraphics[width=\columnwidth]{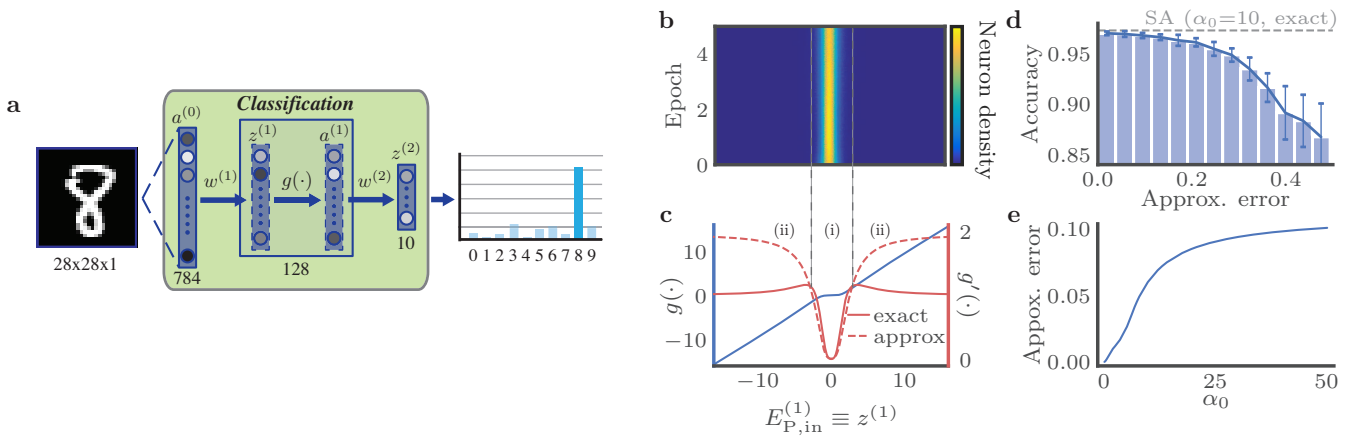}
\caption{\textbf{Effects of imperfect approximation of the activation function derivative. }
\textbf{a,} Feed-forward neural network architecture using a single hidden layer of 128 neurons.
\textbf{b,} Distribution of neuron inputs ($E\nn{1}{\mathrm{P,in}}\equiv z\nn{1}{}$), which is concentrated in the unsaturated region (\romannumeral 1) of the SA activation function, $g(\cdot)$.  As a result, the approximation error in the linear region (\romannumeral 2) is less impactful on the training.
\textbf{c,} The transmission of a SA unit with $\OD{=}10$, along with the exact and (rescaled for easier comparison) optically approximated transmission derivatives.
\textbf{d,} Performance loss associated with approximating activation function derivatives $g'(\cdot)$ with random functions, plotted as a function of the approximation error, for $\OD{=}10$ (see \Cref{Appendix: approximation evaluation} for details).
\textbf{e,} Average  error of the derivative approximation \eqref{eq:sat_abs_forward} as a function of the optical depth of a SA nonlinearity.
}%
\label{fig:2}
\end{figure*}

To investigate our proposed backpropagation scheme, and in particular how our approximated derivatives affect network performance, we consider the canonical ML task of image classification.  Our first set of numerical experiments is to classify images of handwritten digits from 0 to 9.  We use the MNIST~\cite{lecun10} dataset that contains greyscale bitmaps of size $28{\times}28$, which are fed into the input layer of the ONN. The output layer contains 10 neurons whose target values are 0 or 1 dependent on the digit encoded in the bitmap (``one-hot encoding"). In this section we use a network architecture with a single 128-neuron hidden layer as shown in \cref{fig:2}(a). Further details of the networks, training and calculation of the accuracy metric for all experiments presented in this work can be found in \Cref{Appendix: network details}.

Initially, we consider the activation function to be provided by SA with an optical depth of $\OD=10$.  For the chosen network architecture, this provides \SI{97.3\pm0.1}{\percent} classification accuracy after training, with no difference in performance regardless of whether the true derivatives (\cref{eq:SA_grad_exact}) or the optically-obtainable derivative approximations are used.  From \cref{fig:2}(b) we can see that during training the neurons are primarily distributed in the unsaturated region, of the SA activation function.  This is a consequence of the fact that the expressive capacity of neural networks arises from the nonlinearity of its neurons.  Therefore, to train the network, the optically-obtained derivatives need to approximate the exact derivatives (up to a fixed scaling as previously discussed) in only this nonlinear region.

It is interesting to investigate how training is affected by imprecision in the derivatives used. To this end, we evaluate the network performance by replacing the derivative $g'(\cdot)$ with random functions of varying similarity to the true derivative within the nonlinear region (the quantitative measure, $S$, of the similarity is defined in \Cref{Appendix: approximation evaluation}). From \cref{fig:2}(d) we see that the performance appears robust to approximation errors, defined as $1{-}S$, of up to ${\sim} \SI{15}{\percent}$.  We explain this potentially surprising observation by noting that gradient descent will converge even if the update vector deviates from the direction towards the exact minimum of the loss function, so long as this deviation is not too significant.

In the case of SA, \ie{} when the approximate derivatives given by \cref{eq:sat_abs_forward} are used, this error saturates at ${\sim}\SI{10}{\percent}$ for increasing optical depth, see \cref{fig:2}(e), so no significant detrimental effect on the training accuracy can be expected. These results suggest that our scheme would still be effective in a noisy experimental setting and that the approach studied here may function well for a broad range of optical nonlinearities.

\begin{figure*}[t]
\includegraphics{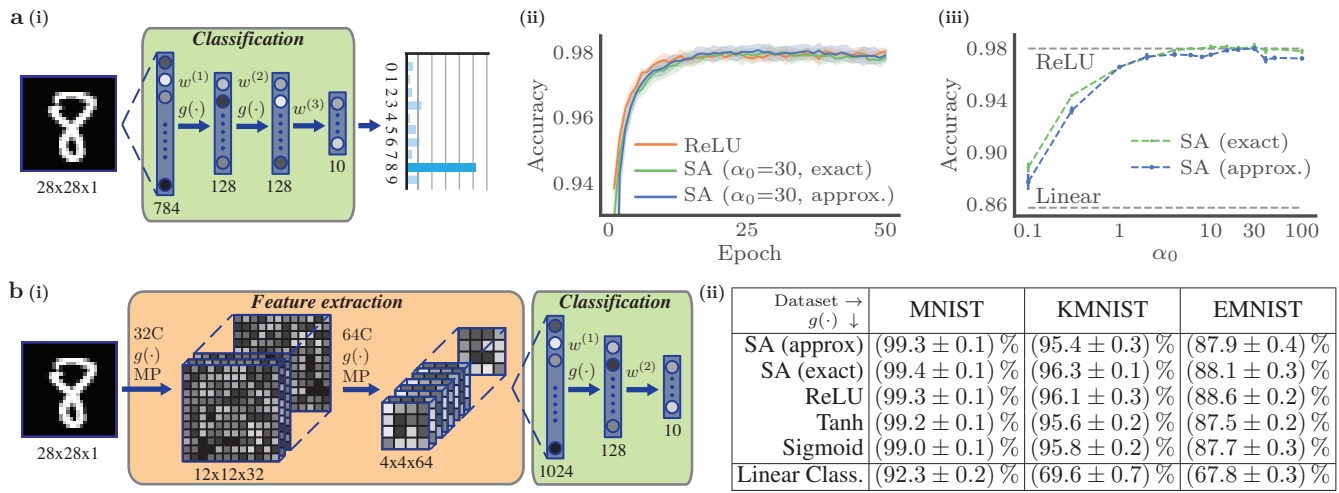}
\caption{\textbf{Performance on image classification.}  \textbf{a,}
(\romannumeral 1)~The fully-connected network architecture.
(\romannumeral 2)~Learning curves for the SA (with either exact or approximated derivatives) and benchmark ReLU networks.  (\romannumeral 3)~The final classification accuracy achieved as a function of the optical depth, $\OD$, of the SA cell.
\textbf{b,}  (\romannumeral 1)~The convolutional network architecture.  Sequential convolution layers of 32 and 64 channels convert a $28{\times}28$ pixel image into a 1024-dimensional feature vector which is then classified (into $N\sub{C}=10$ classes for MNIST and KMNIST, and $N\sub{C}=47$ classes for EMNIST) by fully-connected layers.  Pooling layers are not shown for simplicity.  (\romannumeral 2)~Classification accuracy of convolutional networks when using various activation functions. The same deep network architecture is applied to all datasets, but the SA networks use mean-pooling while the benchmark networks use max-pooling. The last row shows the performance of a simple linear classifier as a baseline.
}%
\label{fig:3}
\end{figure*}

\section{Case study: image classification}

Thus far we have only used a simple network architecture to examine our derivative approximation, however we now consider how ONNs with SA nonlinearities compares to state-of-the-art ANNs.  To do this, we use deeper network architectures for a range of image classification tasks.  To obtain a comparison benchmark, we computationally train ANNs with equivalent architectures using standard best practices.  Concretely, for ANNs we use ReLU (rectified linear unit) activation functions, defined as $g\sub{ReLU}(z) = \max(0,z)$, and the categorical cross-entropy loss function, which is defined as $\Loss {=} {-}\sum_{i}t_{i}\log(p_{i})$ where $p_{i} {=} \exp(z\nn{L}{i}) \Large{/} \sum_k \exp(z\nn{L}{k})$ is the softmax probability distribution of the network output (see \Cref{Appendix: network details} for a discussion of the different choice of loss function for ANNs and ONNs).

To begin, we use a network with two 128-neuron hidden layers as shown in \cref{fig:3}(a)({\romannumeral 1}) and, once again, consider the MNIST dataset.
\Cref{fig:3}(a)({\romannumeral 2}) compares the simulated performance of the optical and benchmark networks. The ReLU-based classifier achieves an accuracy of $\SI{98.0\pm0.2}{\percent}$, which provides an approximate upper bound on the achievable performance of this network architecture for the chosen task~\cite{lecun98}. An optical network with an optical depth of $\OD=30$ exactly matches this level of performance with a $\SI{98.0\pm0.2}{\percent}$ classification accuracy. As an additional benchmark, we train the optical network using the exact derivative, \cref{eq:SA_grad_exact}, of the activation function, obtaining a similar accuracy of  $\SI{98.1\pm0.3}{\percent}$.  The convergence speed to near-optimum performance during training is unchanged across all of these networks.

\Cref{fig:3}(a)({\romannumeral 3}) shows the trained performance of optical networks as a function of the optical depth, which essentially determines the degree of nonlinearity of the transmission function.  As $\OD \rightarrow 0$, our network can only learn linear functions of the input which restricts the classification accuracy to $\SI{85.7\pm0.4}{\percent}$.  For larger optical depths the performance of the network improves, with the strong performance observed at $\OD = 1$ increasing to near optimal levels once $\OD \geq 10$, which is readily obtainable experimentally.  Eventually, for $\OD \geq 30$, we start to see the performance of the approximated derivatives reduced, although high accuracy is still obtained.  This can be attributed to the increasing approximation errors associated with high optical depths (see \cref{fig:2}(e)), which, as previously discussed, accumulate in the deeper network architecture.

To probe the limits of the achievable performance using SA nonlinearities and optical backpropagation, we also consider the more challenging Kuzushiji-MNIST~\cite{clanuwat18} (KMNIST) and Extended-MNIST~\cite{cohen17} (EMNIST) datasets. For these applications we use a deep network architecture with convolutional layers (see \Cref{Appendix: network details} for details), as illustrated in \cref{fig:3}(b)({\romannumeral 1}), which significantly increases the achievable classification accuracy to a level approaching the state-of-the-art.
Whilst not the focus of this work, we emphasise that convolutional operations are readily achievable with optics.  Current research into convolutional ONNs either directly leverages imaging systems~\cite{chang18} or decomposes the required convolution into optical matrix multiplication~\cite{bagherian18, xu19, hamerly19}.

In addition to convolutional layers, convolutional neural networks also contain pooling layers, which locally aggregate neuron activations. The common implementation of these is max-pooling, however this operation does not readily translate to an optical setting. Therefore, for ONNs we deploy mean-pooling, where the activation of neurons is locally averaged, which is a straightforward linear optical operation. In contrast, our benchmark ANNs utilize max-pooling.

\Cref{fig:3}(b)({\romannumeral 2}) compares the obtained performance with SA nonlinearities (with $\OD=10$) to that achieved with benchmark ANNs that use various standard activation functions.  We see an equivalent level of performance, despite the approximation in the backpropagation phase.  This result suggests that all-optical backpropagation can be utilized to train sophisticated networks to state-of-the-art levels of performance.

\section{Discussion}

This work presents an effective and surprisingly simple approach to achieving optical backpropagation through nonlinear units in a neural network -- an outstanding challenge in the pursuit of truly all-optical networks. With our scheme, the information propagates through the network in both directions without interconversion between optical and electronic form. The role of digital electronics is reduced to the preparation of the network input, photodetection and updating the network parameters.  In these elements of the network, the conversion speed is not critical, particularly for large batches of training data.

The scheme is compatible with a variety of ONN platforms; in the Supplemental Material, we discuss  integrated and free-space implementations. We also anticipate that a broader class of nonlinear optical phenomena can be used to implement the activation function.  For example, one could consider directly using saturation of intra-layer amplifiers for this role, circumventing the need for SA units entirely. A preliminary numerical experiment to this effect is discussed in Supplemental Material.  Therefore, as well as presenting a path towards the end-to-end optical training of neural networks, this work sets out an important consideration for nonlinearities in the design of analog neural networks of any nature.

\small

\section*{Acknowledgements}
A.L.'s research is partially supported by Russian Science Foundation (19-71-10092). X.G.\ acknowledges funding from the  University of  Electronic Science and Technology of China.  A.L.\ thanks William Andregg for introducing him to ONNs.

\section*{Author contributions}
The research was conceived by X.G.\ and A.L.\ \ X.G.\ proposed the idea of using SA to implement the activation function. T.D.B.\ and X.G.\ jointly performed the simulations. The manuscript was prepared by T.D.B., X.G.\ and A.L.\ \ All work was done under the supervision of A.L.\ and Z.M.W.

\appendix

\section{Network details}\label{Appendix: network details}
\subsection{Image datasets}

We consider three different datasets, all containing $28{\times}28$ pixel greyscale images: MNIST~\cite{lecun10}, Kuzushiji-MNIST (KMNIST)~\cite{clanuwat18} and Extended-MNIST (EMNIST)~\cite{cohen17}.  MNIST corresponds to handwritten digits from 0 to 9, KMNIST contains 10 classes of handwritten Japanese cursive characters, and we use the EMNIST Balanced dataset, which contains 47 classes of handwritten digits and letters.  MNIST and KMNIST have \num{70000} images in total, split into \num{60000} training and \num{10000} test instances.  EMNIST has \num{131600} images, with \num{112800} (\num{18800}) training (test) instances.  For all datasets, the training and testing sets have all classes equally represented.

\subsection{Network architectures}

The fully-connected network we train to classify MNIST (corresponding to the results in \cref{fig:3}(a)) first unrolls each image into a 784-dimensional input vector, before two 128-neuron hidden layers and a 10-neuron output layer.

The convolutional network depicted in \cref{fig:3}(b)(\romannumeral 1) has two convolutional layers of 32-channel and 64-channels, respectively.  Each layer convolves the input with $5{\times}5$ filters (with a stride of \num{1} and no padding), followed by a nonlinear activation function and finally a pooling operation (with both kernel size and stride of 2).  After the convolutional network, classification is carried out by a fully-connected network with a single 128-neuron hidden layer and $N\sub{C}$-neuron output layer, where $N\sub{C}$ is the number of classes in the target dataset.

Multilayer ONNs are assumed to have the same optical depth of their saturable absorbers in all layers.

\subsection{Network loss function}

As stated in the main text, we train ONNs using the mean-squared-error (MSE) loss function, whereas the ANN baselines use categorical cross-entropy (CCE).  This choice was made as the gradients of MSE loss are readily calculable in an optical setting, whereas the softmax operation in CCE would require offline calculation.  However, our ANNs use CCE as this is the standard choice for classification problems in the deep learning community.  For completeness, we re-trained our ANN baselines for MNIST classification using MSE.  The fully-connected classifier (\cref{fig:3}(a)(\romannumeral 1)) provided a classification accuracy of \SI{98.0\pm0.2}{\percent}, while the convolutional classifier (\cref{fig:3}(a)(\romannumeral 2)), using ReLU nonlinearities, scored \SI{99.5\pm0.1}{\percent}.  In both cases, the performance of MSE is essentially equivalent to that of CCE.

\subsection{Network training}

All networks are trained with a mini-batch size of 64. We used the Adam optimiser with a  learning rate of $5\times10^{-4}$, independent of the optical depth of the SA. 
For each network, the test images of the target dataset are split evenly into a `validation' and `test' set.  After every epoch, the performance of the network is evaluated on the held-out `validation' images.  The best ONN parameters found over training are then used to verify the performance on the `test' set.  Therefore, learning curves showing the performance during training (\ie{} \cref{fig:3}(a)(\romannumeral 2)) are plotted with respect to the `validation' set, with all other reported results corresponding to the `test' set.  The fully-connected networks were trained on MNIST for 50 epochs.  The convolutional networks are trained for 20 epochs when using ReLU, Tanh or Sigmoid nonlinearities, and 40 epochs when using SA nonlinearities.

Training performance is empirically observed to be sensitive to the initialisation of the weights, which we ascribe to the small derivatives away from the nonlinear region of the SA response curve.
For low optical depths, $\OD\leq30$, all layers are initialised as a normal distribution of width 0.1 centred around 0.  For higher optical depths, the weights of the fully-connected ONN shown in \cref{fig:3}(a) are initialised to a double-peaked distribution comprised of two normal distributions of width 0.15 centred at $\pm 0.15$. We do not constrain our weight matrices during training because, as discussed in the main text, conservation of energy can always be satisfied by rescaling the input power or output threshold for the first and last linear transformation and using intra-layer amplifiers in deeper architectures.

For all images, the input is rescaled to be between 0 and 1 (which practically would correspond to $0 \leq E^{(0)}\sub{P,in} \leq 1$) when passing it to an network with computational nonlinearities (\ie{} ReLU, Sigmoid or Tanh).  Due to `absorption' in networks with SA nonlinearities, we empirically observe that rescaling the input data to higher values results in faster convergence when training convolutional networks with multiple hidden layers.  Therefore, the fully connected networks in \cref{fig:3}(a) use inputs between 0 and 1 and the convolutional networks in \cref{fig:3}b use inputs normalised between 0 and 5 (15) for $\OD\leq10$ ($\OD>10$).

\section{Calculation of derivative approximation error}
\label{Appendix: approximation evaluation}

As discussed in the main text, we approximate the true derivatives $g'(\cdot)$ of the activation functions by random functions $f(\cdot)$ to test the effect of the approximation error on training. Here we discuss how these functions are generated and how the similarity measure is defined.

The response of a saturable absorption nonlinearity can be considered in two regimes, nonlinear (unsaturated) and linear (saturated), which are labelled (\romannumeral 1) and (\romannumeral 2) in \cref{fig:SA}, respectively.  During the network training, the neuron input values ($z^{(l)}_{j}$) are primarily distributed in the nonlinear region, as seen in \cref{fig:2}(b) and discussed in the main text.  Therefore, we model the neuron input as a Gaussian distribution within this region,
\begin{equation}
p(z)=\frac{1}{\sqrt{2\pi}\sigma}
\exp\left(-\frac{z^2}{2\sigma^2}\right),
\end{equation}
where $2\sigma$ is the width of region (\romannumeral 1).  We then define the similarity as the reweighted normalised scalar product between the accurate and approximate derivatives,
\begin{align}
\label{Approximation error 2}
S=\frac{|\int f(z)g'(z)p(z)\mathop{d z}|^2}{\int [f(z)]^2 p(z)\mathop{d z}\cdot \int [g'(z)]^2 p(z)\mathop{d z}}.
\end{align}
According to the Cauchy-Schwarz inequality, $S$ is bounded by 1 and therefore so is the average approximation error, $1{-}S$.

To obtain the results in \cref{fig:2}(d-e), we generate 200 random functions for $f$, with different approximation errors.  We first generate an array of pseudo-random numbers ranging from 0 to 1, concatenate it with the flipped array to make them symmetric like the derivative $g'(\cdot)$, and then use shape-preserving interpolation to obtain a smooth and symmetric random function.  The network is then trained once with each of the generated $f$'s.

\section{Code availability}
Source code has been made publicly available at \url{https://github.com/tomdbar/all-optical-neural-networks}.


\end{document}


\title{Supplemental Material: Backpropagation through nonlinear units for all-optical training of neural networks}

\author{Xianxin Guo}
\email{These authors contributed equally to this work.; xianxin.guo@physics.ox.ac.uk}
\affiliation{Institute of Fundamental and Frontier Sciences, University of Electronic Science and Technology of China, Chengdu, Sichuan 610054, China}
\affiliation{University of Oxford, Clarendon Laboratory, Parks Road, Oxford  OX1 3PU, UK}
\affiliation{Institute for Quantum Science and Technology, University of Calgary, Calgary, Canada, T2N 1N4}

\author{Thomas D.\ Barrett}
\email{These authors contributed equally to this work.; thomas.barrett@physics.ox.ac.uk}
\affiliation{University of Oxford, Clarendon Laboratory, Parks Road, Oxford  OX1 3PU, UK}

\author{Zhiming M.\ Wang}
\email{zhmwang@uestc.edu.cn}
\affiliation{Institute of Fundamental and Frontier Sciences, University of Electronic Science and Technology of China, Chengdu, Sichuan 610054, China}

\author{A.\ I.\ Lvovsky}
\email{alex.lvovsky@physics.ox.ac.uk}
\affiliation{University of Oxford, Clarendon Laboratory, Parks Road, Oxford  OX1 3PU, UK}
\affiliation{Russian Quantum Center, Skolkovo, 143025, Moscow, Russia}

\date{\today}

\maketitle

\section{Related works}
Optics provides an attractive platform for the implementation of neural networks, and optical neural networks (ONNs) have been investigated for a long time. In the 1980's to 1990's, most of ONN research focused on the realization of the Hopfield network, a simple type of recurrent neural networks, utilising binary neurons, which serves as an associative memory system~\cite{ONN book-Denz-2013}. This network is designed to store multiple patterns in the fixed weights, and find the best match upon request with a new pattern input -- even when the input only contains partial information of a stored pattern.  In optoelectronic Hopfield networks, the neuron interconnection is realised with optics, for example, via optical vector-matrix multiplication, with the threshold and feedback stages handled with electronically~\cite{Optical Hopfield-Farhat-1985, Optical Hopfield with LCTV-Francis-1990}. In 1987, Abu-Mostafa and his colleagues in Caltech constructed an all-optical Hopfield network~\cite{Caltech Holographic Associative Memory-PSaltis-1987}. They stored four patterns in a hologram, fabricated optical threshold units with nonlinear photorefractive material, and built an optical feedback loop. In that system the weights are encoded as the Bragg grating strengths of the hologram, and neuron interconnection is achieved by the Bragg selectivity. Although Hopfield networks were the first optically constructed and trained neural networks, their application is rather limited, and they represent a fundamentally different paradigm to our focus of optical feedforward networks.

Compared to the Hopfield network, feedforward networks have richer structures to handle different tasks, and they are essential in today's deep learning technology. In 1987, Wagner~\etal{}~\cite{Optical MLP-Wagner-1987} proposed an all-optical multilayer feedforward network that utilised volume holograms for bidirectional neuron interconnection, Fabry-Perot etalons as the nonlinear unit, and optical backpropagation to train the network. Forward- and backward-propagating beams are orthogonally polarised and rotated in the hologram such that they interfere with the correct reference beams for the weighted interconnection in both directions. A bistable Fabry-Perot etalon simulates a sigmoid activation function for forward-propagating strong signals, and roughly approximates the derivative of the sigmoid for backward-propagating weak error fields. To mitigate the significant deviation between the backward etalon response and the exact sigmoid derivative, they suggested insertion of a birefrigent plate or adoption of a dual-etalon scheme where the two etalons are optimized for forward and backward responses respectively. Later they demonstrated proof-of-principle holographic neuron interconnection in forward direction and error-driven update of the hologram~\cite{Holographic interconnection-Wagner-1988}, however the error was calculated digitally.  There has never been a full experimental demonstration of the proposed scheme, largely due to the inherent experimental complexity associated with the polarisation-switching bidirectional volume hologram and the intricate Fabry-Perot etalon design. In our work, whilst we also adopt a similar pump-probe scheme, we aim to simplify the optical backpropagation through the nonlinear unit. Our SA nonlinearity resembles the ReLU activation function, which has replaced sigmoid activations as the standard best practice in modern deep feedforward networks. The SA nonlinearity is also easy to be realised with abundant material choices. Moreover, in our scheme the errors can be backpropagated through the same system without additional experimental complications.

In 1995, Skinner~\etal{}~\cite{ONN proposal-Skinner-1995} proposed a special type of optically trainable feedforward network based on self-lensing media. The wavefront of the forward beam passing through a pattern mask is regarded as a continuum of input neurons, and the electric field coupling during free-space propagation realises the virtual neuron interconnection. The forward beam is then steered by additional ``weight'' beams in a layer of self-lensing media via cross-phase modulation, and propagated towards the detector. In this design, the weight beams can be trained by backpropagating the error beams through the same system. The experimental system was subsequently built and optically trained to classify logic gates~\cite{Optical training-Cabrera-2000}. This is the first, and perhaps only experimental demonstration of an optically-trained feedforward network. However, the network structure differs significantly from that of a regular computational one: there are no discrete neurons, the interconnection cannot be arbitrary since it's achieved via free-space propagation of light, and the nonlinearity is mediated by the ``weights" through the phase.

In recent years, several groups rejoined the effort with advanced optical technologies and realised optical feedforward networks with integrated photonics~\cite{ONN-Englund-2017} and free-space diffraction~\cite{ONN-Ozcan-2018}.  The use of atomic nonlinearities~\cite{ONN-Du-2019} has also been demonstrated. Some of these ONNs were not designed to be trained optically: the training was to be implemented via a digital computer.  In others, brute force training is possible by updating parameters with finite difference method -- perturbing control parameters to directly compute the required gradients~\cite{ONN-Englund-2017} -- yet this is extremely inefficient as compared to the backpropagation algorithm.

A further challenge is that, while the backpropagation algorithm is designed to obtain gradients of the weight matrix for its subsequent updates, in ONNs these weights are accessed not directly, but through a set of actuators --- for example, phase shifters in an integrated interferometer of Ref.~\cite{ONN-Englund-2017}. The mapping to these parameters can be computationally expensive. To tackle with this problem, Hughes~\etal{}~\cite{Optical training-Tyler-2018} recently proposed an \emph{in situ} optical backpropagation and gradient measurement scheme. They showed that gradients of the control parameters in these integrated photonic neural networks can be obtained from a series of forward and backward intensity measurements. Their method may also be extended to other ONN platforms since the derivation starts from the Maxwell equations. However, in their scheme, the nonlinear activation function is assumed to be applied digitally.  The light has to be detected, electronically modulated, and re-injected between each pair of layers in both directions.  Because our scheme considers backpropagation through nonlinear neuron activations, it is complementary to that of Ref.~\cite{Optical training-Tyler-2018} and could be applied in concert with it.

\section{Physical implementation of the optical training scheme}
\subsection{Bidirectional weight matrices}
Bidirectional weighted interconnection of neurons can be experimentally realised with various methods. Here we describe two such methods: integrated photonics and free-space optics.

A real-valued weight matrix can be factorised via singular value decomposition (SVD) into the form $U \Sigma V^{\dag}$, where $U$ and $V$ are unitary matrices and $\Sigma$ is a rectangular diagonal matrix. In optics, it's well known that any unitary matrix can be implemented with a set of Mach-Zehnder interferometers (MZIs) consisting of beam splitters and phase shifters~\cite{OIU-Reck}. The diagonal matrix can be realised with optical attenuators. In integrated photonics, optical interference units (OIUs) with thermo-optical or electro-optical phase shifters together with integrated attenuators can be used to represent the weight matrix. Programmable OIUs with 88 MZIs has been demonstrated~\cite{OIU-Harris}, and a two-layer network with four to five neurons at each layer has been realised. Increasing the network size is a challenging task since the required integrated components in OIUs scales quadratically with the neuron number. To update the weight matrix after obtaining the weight gradients, one still needs to map the new weights to phase shifter settings following the matrix transformation as shown in~\cite{OIU-Reck}. Alternatively, the in situ optical backpropagation scheme~\cite{Optical training-Tyler-2018} can be applied in conjunction with ours to obtain gradients of phase shifter permittivities optically. The update speed of the OIUs is determined by the phase shifter bandwidth, which is usually over \SI{100}{\kHz} for thermo-optical implementations~\cite{TOPS-Harris-2014} and \SI{10}{\GHz} for electro-optical implementations with ferroelectric crystals~\cite{Reed-review-2010}.

\begin{figure}[t]
\includegraphics[width=\textwidth]{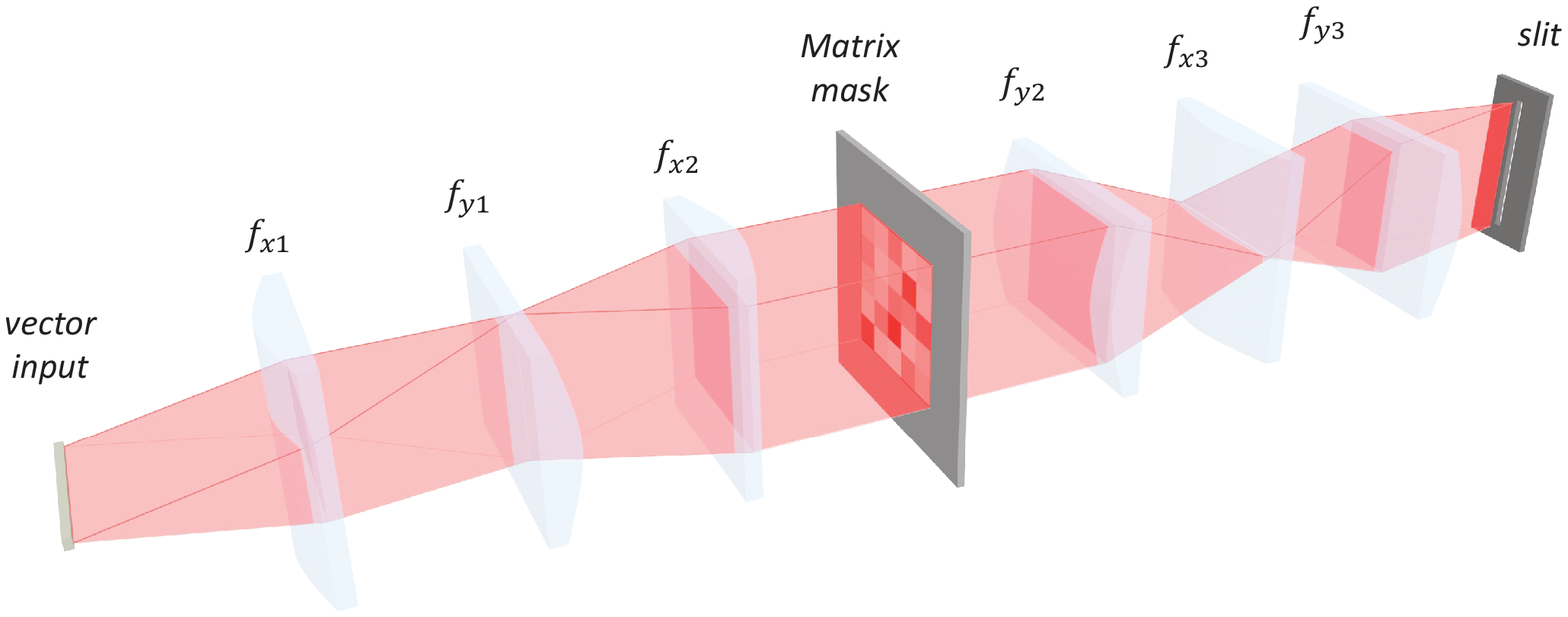}
\caption{\textbf{Free-space coherent optical VMM.} The input vector $a_i$ is prepared as a set of spatial modes distributed horizontally. Each of these modes initially  diverge in the vertical ($y$) dimension until collimation  by a cylindrical lens $f_{y1}$. The vector components are imaged in the horizontal ($x$) dimension by a pair of cylindrical lenses $f_{x1}$ and $f_{x2}$ to the matrix mask plane. In this plane, the vector components are multiplied by the matrix elements $w_{ji}$, so the spatial configuration of the field after the matrix mask is given by $w_{ji}a_i$. A pair of cylindrical lenses $f_{y2}, f_{y3}$ realise 4F imaging of matrix mask plane in the $y$ dimension, and a cylindrical lens $f_{x3}$ realises a Fourier transform in the $x$ dimension. A narrow slit along $y$ is placed at output plane to pass the near-zero spatial frequency components of the Fourier transformed field, corresponding to the summation $\sum_i w_{ji}a_i$. 
}
\label{Fig: VMM}
\end{figure}

In a free-space setting, weighted neuron interconnection can be realised with optical vector-matrix multiplication (VMM). Neuron values are encoded on the electric field of the propagating beam, and real-valued weight matrices can be encoded on liquid-crystal spatial light modulators (LC-SLMs) or digital micromirror devices (DMDs). Precise amplitude and phase control of light can be achieved by modulating the phase grating pattern of the LC-SLMs~\cite{SLM control}. Although DMDs are designed as binary amplitude modulators, multilevel control can be easily achieved by grouping multiple pixels as a unit block.  Taking a block of $10{\times}10$ modulator pixels to represent a neuron/weight block, free-space ONNs with 200-400 neurons per layer can be built with currently available high-resolution LC-SLMs/DMDs. A coherent programmable VMM system can thus be constructed with cylindrical lenses performing $4F$ imaging and Fourier transform~\cite{VMM-Tamura}, as illustrated in \cref{Fig: VMM}. In this setting, only zero spatial frequency components at the output plane carry the correct VMM result, so the output beam has to pass through a narrow optical slit.

To evaluate the power efficiency of the slit, we set the vector and matrix entries to be one so that the output plane shows a sinc spectrum (assuming square aperture of the system), from which we estimate that with average output accuracy of about \SI{95}{\percent}, the power efficiency of the slit is about \SI{50}{\percent}. Higher power efficiency can be obtained at the cost of lower accuracy.

The DMD bandwidth is about \SI{10}{\kHz}, and LC-SLM maximum bandwidth is sub-\si{\kHz}~\cite{SLM review}, hence the update speed of VMM is slower than that of OIU.  An advantage of the free-space implementation, on the other hand, is that each weight in an VMM is independently controlled by a block of pixels on the LC-SLM or DMD.  Therefore, the weight update can be implemented with weight gradients via a calibrated look-up table.

\subsection{Saturable absorber}
Saturable absorption is a common nonlinear optical phenomenon, and there are many different material choices for a saturable absorber in an ONN. An atomic vapor cell or a cold atomic cloud in a magneto-optical trap is a viable option in free-space. Optical depths of $\OD{\gtrsim}10$ can be easily obtained. Although element-wise activation is needed in a neural network, one can accommodate multiple neurons/beams in a single atomic cloud or vapor cell. To prevent the beams from significant divergence inside atomic medium, the Rayleigh length $z\sub{R}=\pi w^2_{0}/\lambda$ should be larger than atomic sample thickness, which is typically on the order of centimeter. Therefore, the beam waist $w_{0}$ in the atomic medium can be about \SI{100}{\micro\meter}, where we take the resonant wavelength of the \Rb{} \DTwo{} line transition. We can accommodate 100 neurons within a sample with a width of \SI{2}{\centi\meter}.

Atomic vapor cells can also be integrated on a silicon chip and coupled to integrated waveguides, as demonstrated in~\cite{Vapor cell on chip-2007, Vapor cell coupled to waveguide-2015}. Optical depth of $\OD{=}1$ to $\OD{=}2$ have been achieved. Other saturable absorbers such as semiconductors or graphene layers featuring low threshold and large modulation bandwidth~\cite{Graphene SA-2009} may also be integrated into nanophotonic circuits~\cite{Graphene in waveguide-2013}. These optical depths allow the ONN to achieve strong performance as demonstrated in the main text.

\subsection{Intra-layer amplification}

In our simulations we do not constrain the weights $w^{(l)}$, but in a realistic passive physical system the weights are bounded because the weighted interconnection must not violate energy conservation. To address this, in a physical implementation, we place an amplifier of gain $A^{(l)}$ in front of a passive weighted interconnection with a matrix $W^{(l)}$, so together they comprise the desired weight matrix  $w^{(l)}=A^{(l)}W^{(l)}$. In this section, we evaluate the necessary gain $A^{(l)}$ of such an amplifier, specialising to the weight matrix $w^{(2)}$ of the simulation in Fig.~4(a) in the main text.

We use two estimates to provide the lower and upper bound of this required gain. For the lower bound, we note that the energy conservation in a passive system implies that
\begin{align}
\sum_{j}(W^{(l)}_{ij})^2 \leq 1 \nonumber \\
\sum_{i}(W^{(l)}_{ij})^2 \leq 1.
\end{align}
In order to satisfy these conditions, the gain $A$ must not be lower than $\max\left(\max_i\sum_{j}(w^{(l)}_{ij})^2,\max_j\sum_{i}(w^{(l)}_{ij})^2\right)$. This bound is plotted in Fig. S2(a). To estimate the upper bound, we take the square of the highest singular value $\Sigma_{\max}$ of the weight matrix $w^{(2)}$. Indeed, if $A\ge\Sigma_{\max}$, then no singular values of  $W^{(2)}$ exceed 1, meaning that this matrix can be implemented as discussed in Sec.~2.1. We plot $\Sigma_{\max}^2$ as a function of the training epoch number in Fig.~S2(b).

Semiconductor optical amplifiers (SOAs) can offer \SI{30}{\dB} amplification with hundreds of \si{\ps} response time, and can be integrated on waveguides~\cite{Semiconductor amplifier-2013}.  From the plot we see that at $\alpha_0 \sim 30$, one stage of power amplification with about 10 dB gain is needed. At lower optical depth, the required gain is generally smaller.

\begin{figure}[t]
\includegraphics[width=0.75\textwidth]{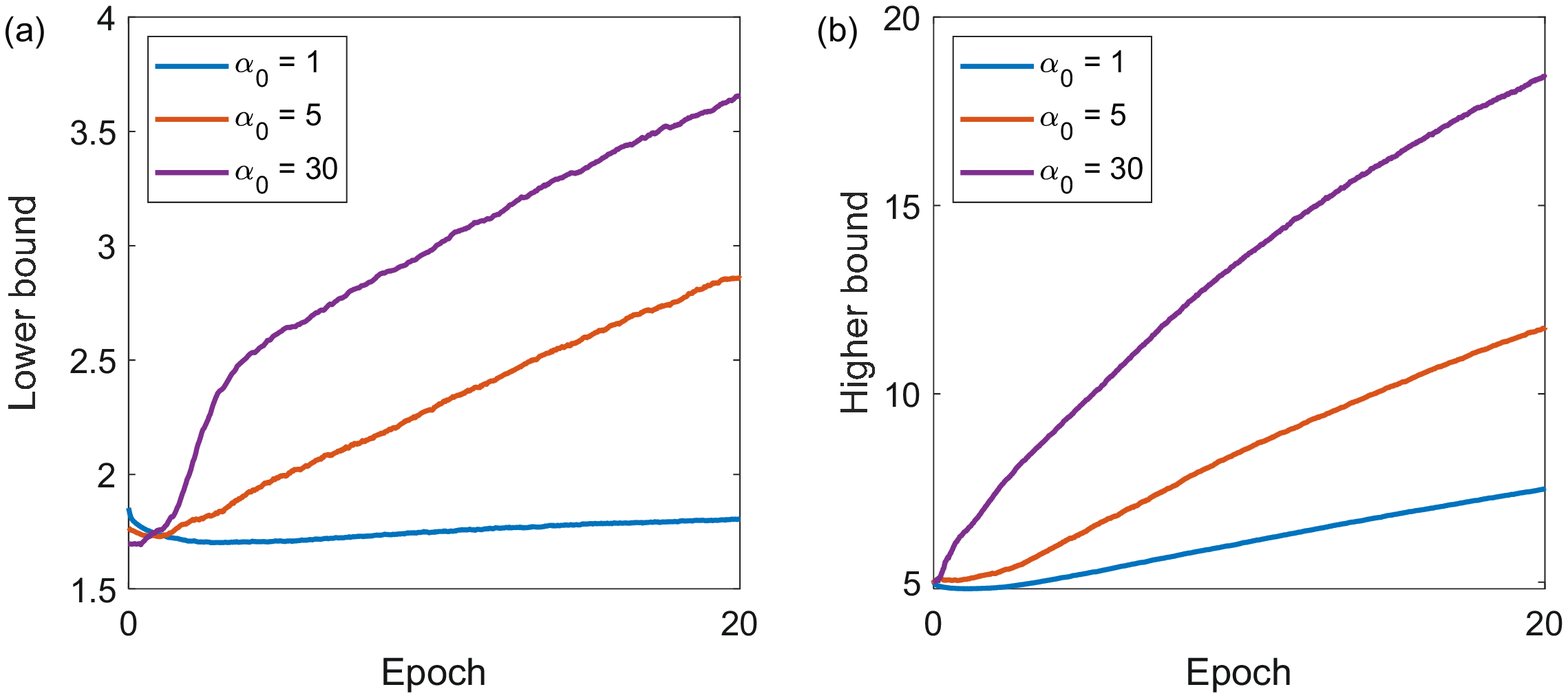}
\caption{Lower (a) and higher (b) bounds on the amplifier gain required for network training with unconstrained weights.
}
\label{Fig: PowerRatio}
\end{figure}

\subsection{Optical power consumption}
The optical power consumption in an ONN depends on the network architecture and implementation details. Therefore, for concreteness, we now consider a fully-connected network with $N= 1000$ units per layer, with SA optical nonlinearities implemented on the \Rb{} \DTwo{} line.
Recalling Fig. 3(b) from the main text, we note that during training the input power to each neuron is typically restricted to the unsaturated
region, (\romannumeral 1), of the nonlinearity response. For the SA nonlinearities we consider, the saturation intensity is given by~\cite{steck}
\begin{equation}
I\sub{sat}=\frac{\hbar \omega \Gamma}{2\sigma_0}=\SI{16.6}{\micro\watt\per\milli\metre\squared},
\end{equation}
where $\Gamma = 2\pi\times\SI{6}{\mega\hertz}$ is the natural linewidth, and $\sigma_0=3\lambda^2/(2\pi)$ is the resonant absorption cross section.  For beams with a waist of $w_{0}=\SI{100}{\micro\meter}$, this corresponds to a saturation power of $P\sub{sat}\approx \SI{500}{\nano\watt}$ per neuron, and total SA input power on the order of \SI{500}{\micro\watt}.

To saturate the SA, the optical pulse needs to be longer than the excited state life time $\Gamma^{-1}=\SI{26}{\nano\second}$. The  energy cost of a single forward pass through the network is then on the order of a fraction of a nanojoule, and the backpropagation energy cost is negligible. Since a single interlayer transition involves a VMM with $N^2$ multiplications, one can estimate the energy cost per floating point operation to be less than a femtojoule. However, these estimates do not include peripheral energy costs in powering and sustaining the instruments and stabilising the system, so the actual power consumption can be expected to be significantly higher.

\section{Optical backpropagation with gain saturation}\label{section-Training with GS}

\begin{figure}[t]
\includegraphics[width=\textwidth]{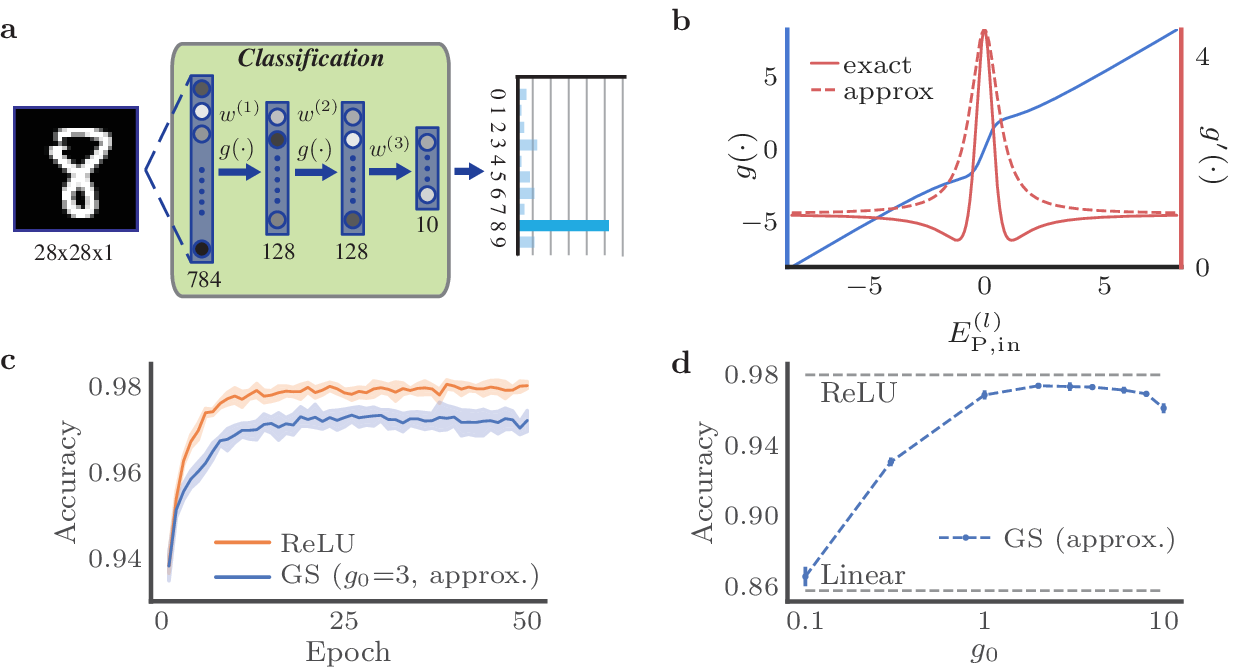}
\caption{\textbf{Optical backpropagation through GS nonlinearity.}
(a) Fully-connected network architecture, which is the same as Fig. 4(a) except for the nonlinearity.
(b) Transmission and transmission derivatives of the GS unit with gain factor $g_{0}=3$.
(c) Learning curves for the GS-based ONN and benchmark ReLU networks. (
(d) The final classification accuracy achieved as a function of the gain. }
\label{Fig: BP with SAMP}
\end{figure}

In optical amplifiers, gain saturation (GS) takes place when a sufficiently high input power depletes the excited state of the gain medium is depleted. In a two-level system, this process can be described similarly to saturable absorption by simply replacing the optical depth term $\alpha_{0}$ in Eq.~(4) of the main text with a positive gain factor $g_{0}$.  The transmission, exact and optically-approximated transmission derivatives are plotted in Fig.~\ref{Fig: BP with SAMP}b with $g_{0}{=}3$. The derivative curves have the inverted shapes of the SA derivative curves, and resemble the sigmoid function derivative.

We use this feature of the GS nonlinearity to implement optical backpropagation. To examine this, we replace the SA nonlinearity with GS nonlinearity in the fully-connected network as shown in Fig. 4(a) of the main text, and repeat the optical training simulation. The MNIST image classification performance is shown in Fig.~\ref{Fig: BP with SAMP}(c, d). High accuracy can be achieved with gain factor as small as 1, and the best result scores \SI{97.3\pm0.1}{\percent} at $g_{0}{=}3$, slightly lower than that of the benchmark ReLU network and SA-based ONN. Since the derivative approximation error of the GS nonlinearity is the same as that of the SA nonlinearity, the performance degradation is mainly attributed to the nonlinearity itself, however, higher performance may be achievable through careful hyperparameter tuning.

